\documentclass[twocolumn,showpacs,superscriptaddress,prl]{revtex4}

\usepackage[latin9]{inputenc}
\usepackage{color}
\usepackage{verbatim}
\usepackage{amsmath}
\usepackage{amssymb}
\usepackage{graphicx}
\usepackage{esint}
\usepackage{hyperref}

\makeatletter
\@ifundefined{textcolor}{}
{%
 \definecolor{BLACK}{gray}{0}
 \definecolor{WHITE}{gray}{1}
 \definecolor{RED}{rgb}{1,0,0}
 \definecolor{GREEN}{rgb}{0,1,0}
 \definecolor{BLUE}{rgb}{0,0,1}
 \definecolor{CYAN}{cmyk}{1,0,0,0}
 \definecolor{MAGENTA}{cmyk}{0,1,0,0}
 \definecolor{YELLOW}{cmyk}{0,0,1,0}
}

\usepackage{mathrsfs}

\draft
\begin{document}

\title{Demonstrating Lattice-Symmetry-Protection in Topological Crystalline Superconductors}

\author{Xiong-Jun Liu}
\affiliation{International Center for Quantum Materials and School of Physics, Peking University, Beijing 100871, China}
\affiliation{Department of Physics, Hong Kong University of Science and Technology, Clear Water Bay, Hong Kong, China}
\affiliation{Collaborative Innovation Center of Quantum Matter, Beijing 100871, China}
\author{James J. He}
\affiliation{Department of Physics, Hong Kong University of Science and Technology, Clear Water Bay, Hong Kong, China}
\author{K. T. Law}
\affiliation{Department of Physics, Hong Kong University of Science and Technology, Clear Water Bay, Hong Kong, China}

\begin{abstract}
We propose to study the lattice-symmetry protection of Majorana zero bound modes in topological crystalline superconductors (SCs). With an induced $s$-wave superconductivity in the $(001)$-surface of the topological crystalline insulator Pb$_{1-x}$Sn$_x$Te, which has a C$_4$ rotational symmetry, we show a new class of 2D topological SC with four Majorana modes obtained in each vortex core, while only two of them are protected by the cyclic symmetry. Furthermore, applying an in-plane external field can break the four-fold symmetry and lifts the Majorana modes to finite energy states in general. Surprisingly, we show that even the C$_4$ symmetry is broken, two Majorana modes are restored exactly one time whenever the in-plane field varies $\pi/2$, i.e. $1/4$-cycle in the direction. This novel phenomenon has a profound connection to the four-fold cyclic symmetry of the original crystalline SC and uniquely demonstrates the lattice-symmetry protection of the Majorana modes. We further generalize these results to the system with generic C$_{2N}$ symmetry, and show that the symmetry class of the topological crystalline SC can be demonstrated by the $2N$ times of restoration of two Majorana modes when the external symmetry-breaking field varies one cycle in direction.
\end{abstract}
\pacs{74.90.+n, 71.10.Pm, 03.65.Vf, 74.78.Na}
\date{\today }
\maketitle

\indent

There has been much attention attracted in realizing topological superconductors (SCs) which support non-Abelian Majorana zero bound modes (MZBMs), driven by both the pursuit of exotic fundamental physics and the potential applications to building fault-tolerant topological quantum computer \cite{Read,Ivanov,Kitaev1,Kitaev2,Fu1,Sato,Sau0,Alicea,Roman0,Oreg,Kouwenhoven,Deng,Das}. While previous studies have been mostly focused on chiral SCs whose topological phases require no symmetry protection, new interests are growing fast in the study of topological SCs which are classified by symmetries~\cite{Ryu,Qi,Teo,Schnyder,Kitaev2009,Beenakker,Law2,Nagaosa,Kane,Teo1,Sato1,Slager,Zhang}. Specifically, in the DIII class topological SCs which are classified by time-reversal symmetry, the Majorana modes come in pairs due to Kramers' theorem~\cite{Qi,Teo,Schnyder,Beenakker,Law2,Nagaosa,Kane}, similar as the boundary modes in the time-reversal invariant topological insulators~\cite{TI}. Due to the protection by time-reversal symmetry, the Majorana Kramers' pairs in DIII class 1D wires are shown to obey non-Abelian statistics~\cite{Liu1,Sato2,Budich,Gaidamauskas}. Moreover, the presence of lattice symmetries~\cite{Fu2} can lead to very wide classes of crystalline topological phases, of which the topological crystalline SCs may support multiple Majorana modes under the protection of lattice symmetries~\cite{Teo1,Sato1,Slager,Zhang}.

While in a topological crystalline SC the MZBMs are protected by lattice symmetries, it is important to ask {\it how to verify such lattice-symmetry protection of Majorana modes in an experiment} (or equivalently, {\it what are the fundamental physics which are uniquely corresponding to such symmetry class})? This question is highly nontrivial, since in general even verifying the existence of multiple Majorana modes in a crystalline SC does not tell the information about what symmetry class the topological superconducting state belongs to. In particular, due to the existing large number of symmetry classes characterized by different space groups~\cite{Slager}, only demonstrating unambiguously the lattice-symmetry protection of Majorana modes can provide unique signatures of the topological crystalline SCs in an experiment.

In this work, we propose to study Majorana modes protected by cyclic symmetries and show how to verify such lattice-symmetry protection of these modes. We base our study on a realistic system by inducing an $s$-wave superconductivity in the $(001)$-surface of the recently observed topological crystalline insulator (TCI) Pb$_{1-x}$Sn$_x$Te which has attracted attention~\cite{Fu3,Xu}. The predicted 2D topological crystalline SC is of cyclic symmetries which are shown to protect two Majorana modes in each vortex core. Furthermore, we unveil a novel phenomenon that the response of the Majorana modes to a symmetry-breaking field exhibits the periodicity which intrinsically corresponds to the original cyclic symmetry, providing an unambiguous verification of the symmetry class of the original system and the lattice-symmetry protection of the MZBMs. We further generalize these results to the system with generic C$_{2N}$ symmetry.

\begin{figure}[t]
\includegraphics[width=0.7\columnwidth]{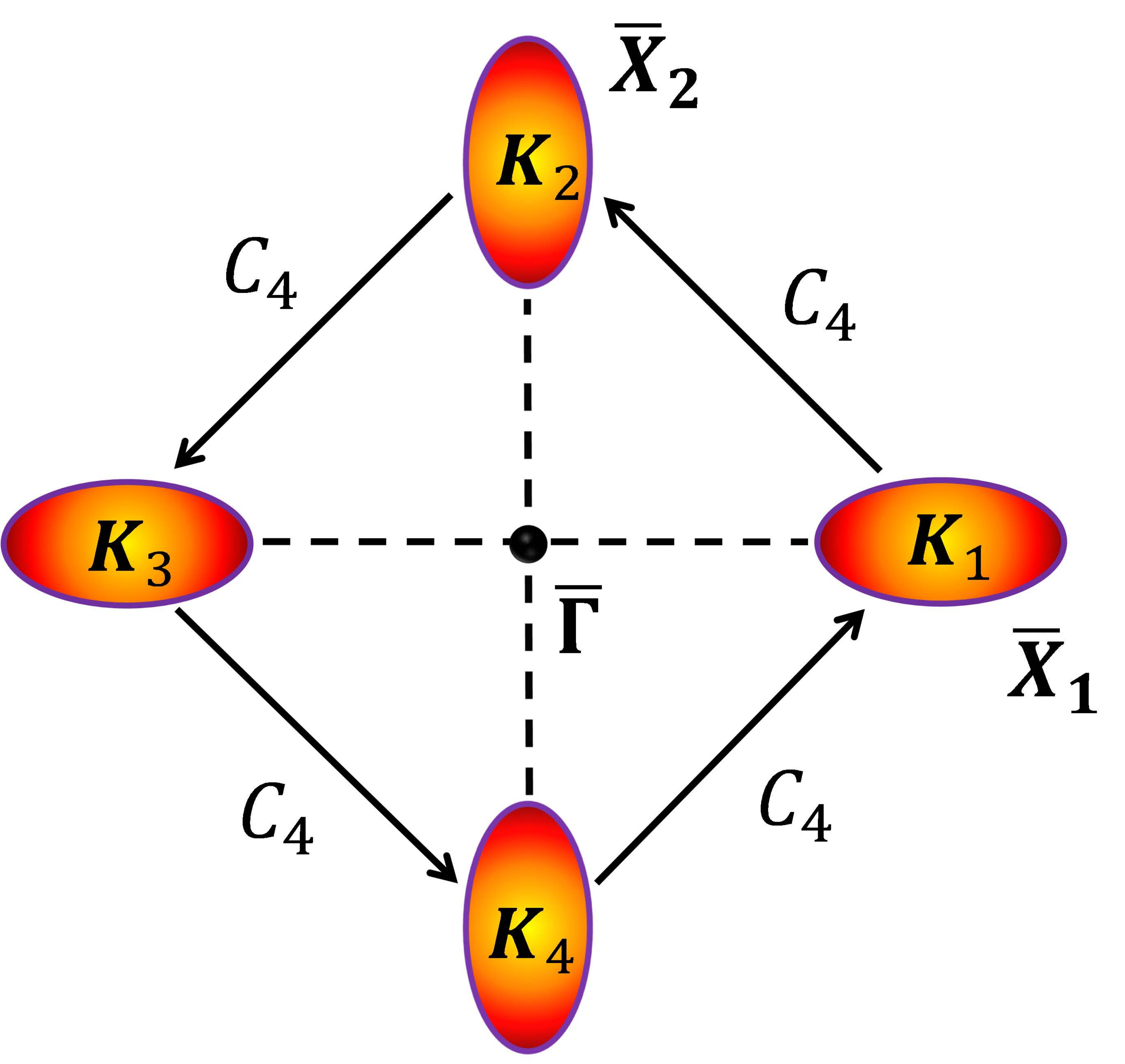}\caption{(Color online) Sketch of four surface Dirac cones $\bold K_j$ ($j=1,2,3,4$) located in $\bar\Gamma\bar X_1$ and $\bar\Gamma\bar X_2$ lines in the (001)-surface of Pb$_{1-x}$Sn$_x$Te with anisotropic Fermi velocities. The arrows imply that under the $C_4$ four-fold rotation the Dirac cones transforms as $\bold K_j\rightarrow\bold K_{j+1}$ and $\bold K_4\rightarrow\bold K_{1}$.}
\label{Dirac}
\end{figure}
The surface states of the TCI Pb$_{1-x}$Sn$_x$Te can be described with even number of Dirac cones protected by lattice symmetries. In particular, for the (001)-surface, the low energy physics of the surface states are captured by four Dirac cones respectively centered at four Dirac points $\bold K_j$ ($j=1,2,3,4$), with $\bold K_1=-\bold K_3$ and $\bold K_2=-\bold K_4$, as sketched in Fig.~\ref{Dirac}~\cite{Fu3,Xu}. Inherited from the TCI Pb$_{1-x}$Sn$_x$Te, the surface Hamiltonian preserves mirror symmetries along $\bar\Gamma\bar X_1$ and $\bar\Gamma\bar X_2$ directions. Furthermore, the four Dirac cones are transformed according to $\bold K_j\rightarrow\bold K_{j+1}$ and $\bold K_4\rightarrow \bold K_1$ under the $C_4$ four-fold rotation (Fig.~\ref{Dirac}). Using $k_x$ and $k_y$ to form a right-handed coordinate centered at $\bold K_1$ on the surface, the surface Dirac Hamiltonian around $\bold K_1$ reads
\begin{eqnarray}
H_{\rm sf}^{(1)}=v_\perp k_x\sigma_y-v_\parallel k_y\sigma_x,
\end{eqnarray}
where $v_\perp$ ($v_\parallel$) represents the anisotropic Fermi velocity in the $k_x$ ($k_y$) direction, and $\sigma_{x,y}$ are Pauli matrices in the spin space. For the spin system, the C$_4$ symmetry is defined by $C_4(\pi/2)=\hat R_z(\pi/2)\exp{(-i\pi \sigma_z/4)}$, with which the other three Dirac Hamiltonians can be obtained. Here $\hat R_z(\pi/2)$ is the 2D rotation transformation on Bravais lattice vector as $(R_x,R_y)\rightarrow(R_y,-R_x)$ \cite{Fu3}. In the presence of C$_4$ symmetry, the four Dirac cones are decoupled from each other and the surface states are topologically stable.

Now we consider an induced $s$-wave superconductivity in the surface states through the $s$-wave superconducting proximity effect~\cite{Fu1,Sato,Sau0,Alicea,Roman0,Oreg,Kouwenhoven,Deng,Das}. In particular, a realistic way to induce superconductivity in the (001)-surface of the TCI material Pb$_{1-x}$Sn$_x$Te is to grow a thin-film TCI from the (001)-direction~\cite{Fu4} on an $s$-wave superconductor substrate.
Note that the SC pairing occurs between two states related by time-reversal symmetry, which implies that the pairing only occurs between the states respectively belonging to the Dirac cones around $\bold K_1$ ($\bold K_2$) and $\bold K_3$ ($\bold K_4$). Denoting by $\Delta_s$ the $s$-wave SC order, we obtain the total Hamiltonian
\begin{eqnarray}\label{eqn:surface2}
H&=&\sum_{\bold k,j=1,3}(-iv_\perp k_xc_{j\uparrow,\bold k}^\dag c_{j\downarrow,\bold k}-v_\parallel k_yc_{j\uparrow,\bold k}^\dag c_{j\downarrow,\bold k}+{\rm H.c.})\nonumber\\
&+&\sum_{\bold k}(\Delta_sc_{1\uparrow,\bold k} c_{3,\downarrow,-\bold k}-\Delta_sc_{1\downarrow,\bold k} c_{3,\uparrow,-\bold k}+{\rm H.c.})\nonumber\\
&-&\sum_{\bold k,s,j=1,3}\mu c_{js,\bold k}^\dag c_{js,\bold k}+\biggr\{C_4 \ {\rm terms}\biggr\},
\end{eqnarray}
where $\mu$ is the chemical potential and $c_{j,s}$ ($c^\dag_{j,s}$) is the annihilation (creation) operator of a surface electron from the $j$-th Dirac cone and in the spin state $s=\uparrow,\downarrow$. The last term represents the C$_4$ transformation on all the former terms.
The Hamiltonian~\eqref{eqn:surface2} can be block diagonalized with a set of new bases $c_{a_\pm s,\bold k}=\frac{1}{\sqrt{2}}(c_{1s,\bold k}\pm c_{3s,\bold k})$ and $c_{b_\pm s,\bold k}=\frac{1}{\sqrt{2}}(c_{2s,\bold k}\pm c_{4s,\bold k})$ with $s=\uparrow,\downarrow$.
In terms of new bases we obtain $H=\sum_{\bold k,\eta=\pm}\bigr[H_{a_\eta}(\bold k)+H_{b_\eta}(\bold k)\bigr]$, where
\begin{eqnarray}
H_{a_\pm/b_\pm}(\bold k)&=&\bigr[(-iv_{\perp/\parallel} k_x-v_{\parallel/\perp} k_y)c_{a_\pm/b_\pm,\uparrow,\bold k}^\dag c_{a_\pm/b_\pm,\downarrow,\bold k}\nonumber\\
&\pm&\Delta_sc_{a_\pm/b_\pm,\uparrow,\bold k} c_{a_\pm/b_\pm,\downarrow,-\bold k}+{\rm H.c.}\bigr]\nonumber\\
&-&(\mu/2)\sum_{s}c_{a_\pm/b_\pm,s,\bold k}^\dag c_{a_\pm/b_\pm s,\bold k}.\label{eqn:surface3}
\end{eqnarray}
The $s$-wave order may change the symmetry respected by the total Hamiltonian. In particular, to study Majorana modes we shall consider a vortex profile of the $s$-wave order parameter $\Delta_s=\Delta_0(r)e^{in\theta(\bold r)}$, with $n$ the winding number, which breaks mirror symmetry. However, one can verify that the Hamiltonian is still invariant under a new C$_4$ rotational transformation defined as $C^{(n)}_4(\pi/2)=\exp(in\pi/4)C_4(\pi/2)$. Indeed, this result can be confirmed with the results that under the C$_4$ transformation the superconducting vortex satisfies $\hat R_z^{-1}(\pi/2)\Delta_s(\bold r)\hat R_z(\pi/2)=i^n\Delta_s(\bold r)$ and the bases $(c_{a_\pm s,\bold k},c_{b_\pm s,\bold k})\rightarrow e^{i(n-l)\pi/4}(c_{b_\pm s,\bold k'},-c_{a_\pm s,\bold k'})$,
where the momentum $\bold k'=(-k_y,k_x)$ and $l=1(-1)$ for $s=\uparrow(\downarrow)$. From the Eq.~\eqref{eqn:surface3} we find that the Hamiltonian $H$ describes four decoupled Dirac cones with the induced s-wave orders $\pm\Delta_s$, respectively, which render four copies of chiral $p$-wave topological SC with the effective SC pairing $\propto p_x+ip_y$~\cite{Fu1}. However, the bases $c_{a_\pm}$ and $c_{b_\pm}$ are connected to each other by the C$_4$ transformation, which distinguishes our system from a trivial four-copy version of chiral $p$-wave SCs, and can lead to new physics in the present 2D topological crystalline SC.

In a chiral $p_x+ip_y$ SC, each vortex with an odd vorticity can bind one Majorana mode~\cite{Read}. The analytic wave functions of the MZBMs can be obtained by assuming $v_\perp=v_\parallel=v_F$. While the actual bound state wave functions should be obtained by varying continuously the Fermi velocities to be anisotropic, the topology and the symmetry protection of the MZBMs obtained below are independent of the anisotropy since the bulk gap is not closed. Noting that the subspace of $h_{a_\pm}$ is related to that of $h_{b_\pm}$ by the C$_4$ transformation, the solutions to $h_{b_\pm}$ can be obtained by the C$_4$ operation on those to $h_{a_\pm}$. In the Nambu space $\psi_a(\bold r)=[\psi^{(+)},\psi^{(-)}]^T$ with $\psi^{(\pm)}=\bigr[c_{\pm\uparrow}(\bold r),c_{\pm\downarrow}(\bold r),c^\dag_{\pm\downarrow}(\bold r),-c^\dag_{\pm\uparrow}(\bold r)\bigr]^T$, we write down the Hamiltonian by
\begin{eqnarray}
H_a=\int d^2\bold r\psi^\dag(\bold r){\cal H}_a(\bold r)\psi(\bold r),
\end{eqnarray}
where ${\cal H}_a={\cal H}_{a_+}(\bold r)\oplus{\cal H}_{a_-}(\bold r)$ and ${\cal H}_{a_\pm}\bold r)=v_F(k_xs_y-k_ys_x)\tau_z-\mu\tau_z\pm\Delta_0(r)\big[\cos n\theta(\bold r)\tau_x+\sin n\theta(\bold r)\tau_y\bigr]$.
For an odd winding number $n=2m+1$, the Majorana modes for ${\cal H}_{a_\eta}(\bold r)$ with $\eta=\pm$ are given by (see Appendix for details)
\begin{eqnarray}\label{eqn:MF1}
\gamma_a^{(\pm)}&=&i^{\frac{1\pm1}{2}}\int d^2\bold ru_0(r)\biggr\{J_m(\frac{\mu}{v_F}r)\bigr[c_{a_\pm\uparrow}(\bold r)\mp c^\dag_{a_\pm\uparrow}(\bold r)\bigr]\nonumber\\
&+&J_{m+1}(\frac{\mu}{v_F} r)\bigr[c_{a_\pm\downarrow}(\bold r)e^{i\theta(\bold r)}\mp c^\dag_{a_\pm\downarrow}(\bold r)e^{-i\theta(\bold r)}\bigr]\biggr\}e^{im\theta},\nonumber
\end{eqnarray}
where $J_m(\frac{\mu}{v_F}r)$ is the $m$-th Bessel function and $u_0(r)=e^{-\int_0^rdr'\Delta_0(r')dr'}$. The other two MZBSs for $H_{b_\pm}$ are obtained by C$_4$ rotation and their relations read
\begin{eqnarray}\label{eqn:MF3}
\gamma^{(\eta)}_b=C_{4}^{(n)^{-1}}\gamma^{(\eta)}_aC_4^{(n)},\ C_{4}^{(n)^{-1}}\gamma^{(\eta)}_bC_4^{(n)}=\eta\gamma^{(\eta)}_a.
\end{eqnarray}
These relations can be reorganized in a more transparent way by redefining that $\gamma_{1,2}=(\gamma^{(+)}_{a,b}+\gamma^{(-)}_{a,b})/\sqrt{2}$ and $\gamma_{3,4}=(\gamma^{(+)}_{a,b}-\gamma^{(-)}_{a,b})/\sqrt{2}$. Then we have
\begin{eqnarray}\label{eqn:MF4}
\gamma_{j+1}=C_{4}^{(n)^{-1}}\gamma_{j}C_4^{(n)},\ \gamma_1=C_{4}^{(n)^{-1}}\gamma_{4}C_4^{(n)}.
\end{eqnarray}
The Eq.~\eqref{eqn:MF4} reflects the cyclic properties of the four Majorana modes, which determine that the only possible perturbation respecting C$_4$ symmetry takes the form $V_{\rm pert}=\Gamma_0(i\gamma_1\gamma_2+i\gamma_2\gamma_3+i\gamma_3\gamma_4+i\gamma_4\gamma_1)=i2\Gamma_0\gamma^{(-)}_a\gamma^{(-)}_b$. Thus an infinitesimal perturbation without breaking C$_4$ symmetry can gap out $\gamma^{(-)}_{a,b}$. However, the remaining two Majorana modes $\gamma^{(+)}_a$ and $\gamma^{(+)}_b$ are protected by the C$_4$ symmetry. Therefore, the four-fold cyclic group protects only two MZBMs. We note that in a realistic topological crystalline SC the defects or disorders generically break the C$_4$-rotational symmetry. However, the hybridization between $\gamma_{a}^{(+)}$ and $\gamma^{(+)}_b$ vanishes if  within their localization length the disorder respects the four-fold symmetry by average. Under this condition the C$_4$ symmetry by average provides the protection of the two Majorana modes. In the Appendix we have confirmed this result by numerical simulation.

The existence of Majorana zero modes can lead to zero bias peak in the tunneling transport from a normal lead to the vortex core of the SC~\cite{Kouwenhoven,Deng,Das,Law,Flensberg,Wimmer,Liu2,Sun2014}. Without breaking the C$_4$ symmetry, due to the two symmetry protected MZBMs, the height of zero bias peak is double of that induced by a single MZBM under the same conditions. In particular, at zero temperature, the zero bias peak is of height $4e^2/h$~\cite{Law2}. These results may reflect the number of Majoranas localized in each vortex core.

After showing the existence of two protected Majorana modes, it is important to study how to demonstrate in an experiment the protection by the lattice symmetry. For this we consider an external field to induce lattice-symmetry-breaking terms in the Hamiltonian. In particular,
with the consideration of a distortion by strain or an in-plane magnetic field~\cite{Fu3} (or an electric field by a gate~\cite{Fu4}) one can break the four-fold symmetry and introduce mass terms for the four Dirac cones in the form $m_j=\alpha(\bold u\times\bold K_j)\cdot\hat e_z$ or $m_j=\alpha(\bold u\cdot\bold K_j)$, where $\bold u$ represents the direction of the displacement vector due to distortion or the applied external field. By a straightforward derivative one can find that the mass terms generate couplings between the MZBMs $\gamma^{(\pm)}_a$ and $\gamma^{(\pm)}_b$, and the total coupling Hamiltonian takes the following form
\begin{eqnarray}\label{eqn:coupling2}
V_{\rm c}(\phi)&=&i2\Gamma_0\gamma^{(-)}_a\gamma^{(-)}_b+i2\Gamma_1\sin(2\phi+\beta)\gamma^{(+)}_a\gamma^{(+)}_b+\nonumber\\
&+&i2m_0\cos\phi\gamma^{(+)}_a\gamma^{(-)}_a+i2m_0\sin\phi\gamma^{(+)}_b\gamma^{(-)}_b,
\end{eqnarray}
where $\phi$ represents the angle of the symmetry-breaking field, and $m_0$ is the amplitude of the induced coupling.
It is noteworthy that since the C$_4$ symmetry is broken, an additional $\Gamma_1$-coupling term between $\gamma^{(+)}_a$ and $\gamma^{(+)}_b$ is now allowed. On the other hand, the variation of $\phi$ by $\pi/2$ is equivalent to do a $C_4^{(n)}(\pi/2)$ rotation, which transforms $\gamma^{(+)}_a\leftrightarrow\gamma^{(+)}_b$, and change the sign of the coupling coefficient. Therefore, this coupling term must be proportional to $\sin(2\phi+\beta)$, with $\beta$ an arbitrary constant which can be material-dependent. We note that in general the perturbation by an in-plane vector- or pseudovector-type field always induces the symmetry-breaking couplings as given in Eq.~\eqref{eqn:coupling2}, while we shall show later that the physics studied below are independent of the details of the coupling Hamiltonian $V_{\rm c}(\phi)$.

\begin{figure}[t]
\includegraphics[width=1\columnwidth]{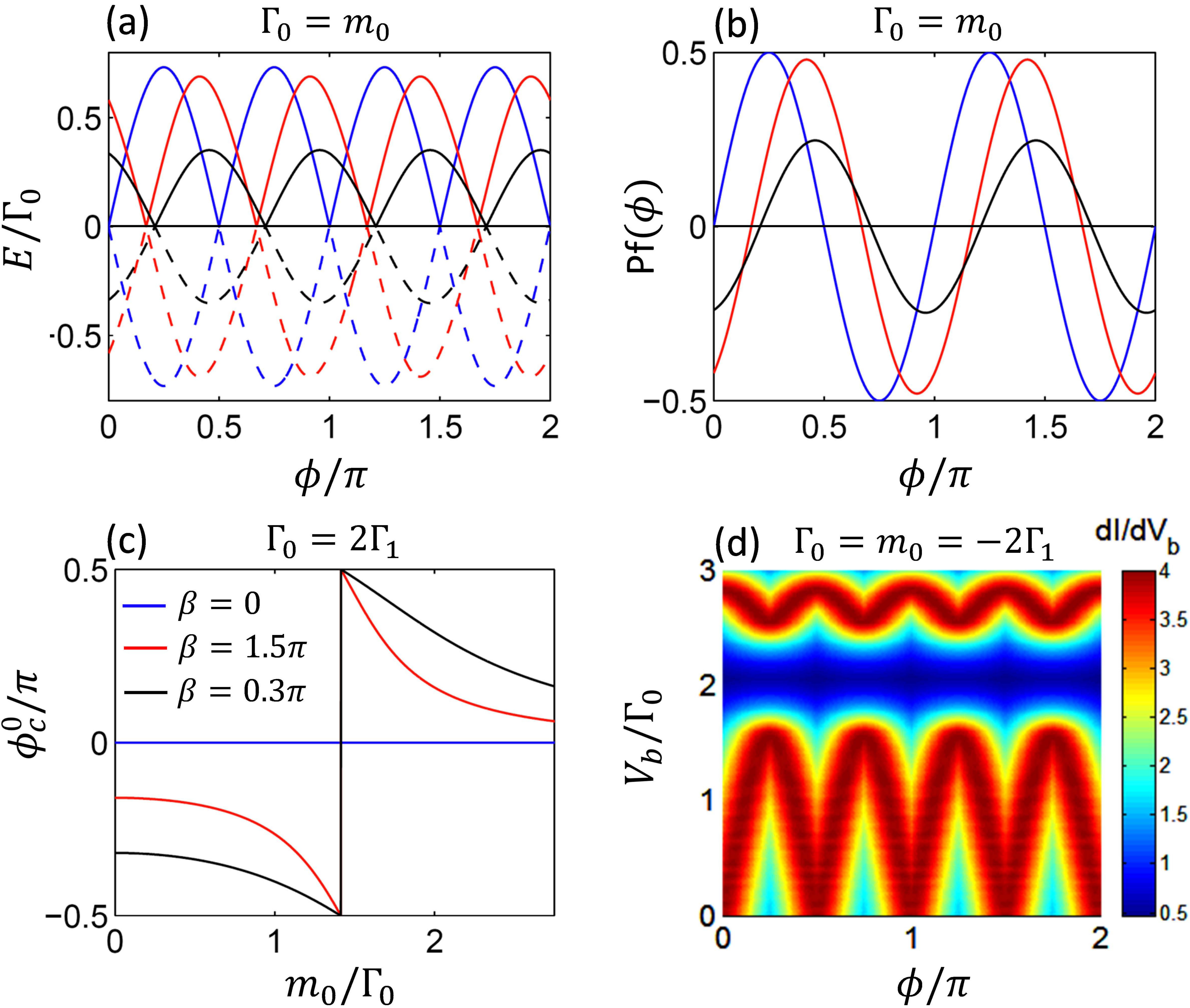}\caption{(Color online) (a) Spectra of the lowest two Andreev bound states by breaking C$_4$ symmetry and (b) Pfaffian versus $\phi$, with the parameters $\Gamma_1=0$ (blue curves); $\Gamma_1=0.5\Gamma_0, \beta=0.5\pi$ (black curves); and $\Gamma_1=0.5\Gamma_0,\beta=1.0\pi$ (red curves), respectively. (c) Critical angle $\phi_c^0$ versus $m_0$ for different magnitudes of $\beta$. (d) The differential tunneling conductance in unit of $e^2/h$ at zero temperature, with $\Gamma_0=m_0=2\Gamma_1$, $\beta=0$, and the tunneling energy chosen to be $0.2\Gamma_0$. The upper bright band represents the higher Andreev bound state spectrum. Only the results with bias $V_b\geq0$ are shown here.}
\label{C4symmetry}
\end{figure}
The spectra of the coupling Hamiltonian~\eqref{eqn:coupling2} can be numerically solved, with two positive and two negative eigenvalues obtained for a fixed $\phi$. It is interesting that for any input parameters of $\Gamma_{0,1},\beta,$ and $m_0$, as long as $m_0^2\neq2\Gamma_0\Gamma_1$ (or $\beta\neq0$), one can always find that spectra for the two lowest Andreev Bound states (ABSs) cross zero energy four times when the angle $\phi$ varies a cycle from $0$ to $2\pi$. Several cases are shown in Fig.~\ref{C4symmetry} (a). This phenomenon has a profound reason which is intrinsically related to the C$_4$ symmetry of the original system. Actually, when $\phi$ advances $\pi/2$, we find that the coupling Hamiltonian
\begin{eqnarray}
V_{\rm c}(\phi+\pi/2)&=&i2\Gamma_0\gamma^{(-)}_a\gamma^{(-)}_b-2\Gamma_1\sin(2\phi+\beta)\gamma^{(+)}_a\gamma^{(+)}_b\nonumber\\
&-&i2m_0\sin\phi\gamma^{(+)}_a\gamma^{(-)}_a +i2m_0\cos\phi\gamma^{(+)}_b\gamma^{(-)}_b,\nonumber\\
\end{eqnarray}
which can return to $V_{\rm c}(\phi)$ by C$_4$ rotation that $\gamma_a^{(\pm)}\rightarrow\gamma_b^{(\pm)}$ and $\gamma_b^{(\pm)}\rightarrow\pm\gamma_a^{(\pm)}$. On the other hand, note that the four MZBMs can form two complex fermions (ABSs) as $f_{a,b}=\gamma_{a,b}^{(+)}+i\gamma_{a,b}^{(-)}$. Then the variation of $\pi/2$ in $\phi$ equivalently transforms $f_a\rightarrow f_b$ and $f_b\rightarrow f_a^\dag$ according to Eq.~\eqref{eqn:MF3}, which implies that the ground state of the system changes fermion parity. This result can be quantitatively confirmed by calculating the Pfaffian~\cite{Kitaev1} for the coupling Hamitonian $V_{\rm c}(\phi)$, which can be written as
\begin{eqnarray}
V_{\rm c}(\phi)=i\sum_{\alpha,\alpha',\eta,\eta'}\gamma_{\alpha}^{(\eta)}{\cal V}_c(\phi)\gamma_{\alpha'}^{(\eta')},
\end{eqnarray}
with $\alpha,\alpha'=a,b$ and $\eta,\eta'=\pm$.
The matrix ${\cal V}_c(\phi)$ is skew-symmetric, with the Pfaffian satisfying
\begin{eqnarray}\label{eqn:Pfaffian}
\frac{{\rm Pf}\bigr[{\cal V}_c(\phi)\bigr]}{{\rm det}\bigr[{\cal V}_c(\phi)\bigr]^{\frac{1}{2}}}={\rm sgn}\biggr[\sin(2\phi)-\frac{2\Gamma_0\Gamma_1}{m_0^2}\sin(2\phi+\beta)\biggr],
\end{eqnarray}
which reverses sign when varying $\phi$ by $\pi/2$, with $m_0^2\neq2\Gamma_0\Gamma_1$ (or $\beta\neq0$) [Fig.~\ref{C4symmetry} (b)]. Therefore, there must be one zero-energy crossing in the ABS spectra during this variation. Finally, when the angle $\phi$ advances $2\pi$, we always get four crossings for the lowest two ABSs, and the four angles of the crossing points are given by
\begin{eqnarray}\label{eqn:angle}
\phi_c^l=\frac{1}{2}\tan^{-1}\biggr[\frac{2\Gamma_0\Gamma_1}{m_0^2-2\Gamma_0\Gamma_1}\tan\beta\biggr]+\frac{l}{2}\pi, \ l=0,...,3,
\end{eqnarray}
at which points two Majorana modes are obtained [Fig.~\ref{C4symmetry} (c)]. The cyclic zero-energy crossings can be measured by the differential tunneling conductance, as shown in Fig.~\ref{C4symmetry} (d). This novel phenomenon can be adopted to demonstrate the symmetry-protection for MZBMs by C$_4$ group of the original system.

Finally we show how to generalize the above results for a system with C$_4$ symmetry to the system with generic C$_{2N}$ symmetry, in which case the surface has $2N$ Dirac cones related by the $2N$-fold rotational symmetry. Note that the four MZBMs $\gamma^{(\pm)}_{a,b}$ consist of a 4D reducible representation of C$_{4}$ group. In the diagonal bases given by $\gamma_\pm=\gamma^{(+)}_a\pm\gamma^{(+)}_b$, $f=\gamma^{(-)}_a+i\gamma^{(-)}_b$ and $f^\dag=\gamma^{(-)}_a-i\gamma^{(-)}_b$, the group element can be represented as $M[C_4^{(n)}]=[1,-1,-i,i]^{\rm diag}$. Then each of the two Majorana bases and two complex fermion bases consists of a 1D irreducible representation. For a unitary group, it can be shown that only the Majorana bases ($\gamma^{(+)}_{a,b}$) which consist of independent 1D irreducible representations are protected by the symmetry group, while the complex fermion bases are not~\cite{note1}. This result provides an alterative interpretation of the protection of two MZBMs in the C$_4$-symmetric system.
Similarly, for a system with C$_{2N}$ symmetry, from the $2N$ MZBMs we can obtain $2N$ independent 1D irreducible representations of the C$_{2N}$ group, with the representation matrix given by $M[C_{2N}^{(n)}]=[1,e^{i\pi/N},e^{i2\pi/N},...,-1,...,e^{i(2N-1)\pi/N}]^{\rm diag}$. Again only the two real eigenvalues $\pm1$ correspond to the Majorana bases (denoted by $\gamma_\pm$) for the representation space, which are protected by the C$_{2N}$ symmetry. The remaining bases are complex fermion modes and can be gapped out without breaking C$_{2N}$ symmetry. With this general proof one can also know that a C$_{2N+1}$ symmetry cannot protect any Majorana mode.

For the generic C$_{2N}$-symmetric system, applying an external field can gap out all the Majorana modes but restore two of them $2N$ times when the field varies one cycle in the direction. Actually, when the angle $\phi$ of the field advances $\pi/N$, the new coupling Hamiltonian $V_{\rm couple}(\phi+\pi/N)$ should generically return to $V_{\rm couple}(\phi)$ by a C$_{2N}$ rotation: $V_{\rm c}(\phi)=C_{2N}^{(n)^{-1}}V_{\rm c}(\phi+\pi/N)C_{2N}^{(n)}$. The C$_{2N}$ rotation transforms the complex fermion mode $f=\gamma_++i\gamma_-$ through $f\rightarrow f^\dag$ and $f^\dag\rightarrow f$ according to the eigenvalues of the representation matrix $M[C_{2N}^{(n)}]$. Therefore the ground state changes fermion parity and one level crossing occurs. The generality of this result can be understood in an intuitive picture. Namely, in the coupling Hamiltonian $V_{\rm c}(\phi)$ the C$_{2N}$ symmetry is broken for any fixed $\phi$. However, if the external field co-rotate with the C$_{2N}$ transformation, the entire system including the topological crystalline SC and the field is C$_{2N}$-invariant. This property ensures that the $2N$-time restoration is independent of the details of the coupling Hamiltonian and can be a unique signature of the protection of Majorana modes by the C$_{2N}$ symmetry.

Before conclusion we provide further discussions on the realistic experimental observations. Note that the interface between a TCI and a substrate superconductor is not easily accessible, and a realistic measurement can be carried out when Majorana zero modes also exist in the vortex core of the top surface. This might be achieved by considering a TCI thin film with a few atomic layers so that the superconductivity can be induced in both the top surface and the TCI/superconductor interface. On the other hand, a sufficient minigap in the vortex core, given by $\sim|\Delta_s|^2/E_F$ with $E_F$ the Fermi energy away from the Dirac point, is necessary to distinguish the zero bias peak of Majorana zero modes from the contribution by low-energy ABSs. Besides tuning $E_F$ close to the Dirac point, another promising way to enhance minigap is to apply a d-wave high Tc superconductor substrate, which does not change the above predicted results, but can greatly enhance the proximity induced superconductivity in the top surface. Based on the recent experimental results~\cite{Zhou2013}, a full superconducting gap in the order of $10$meV is obtained in the surface states of a topological insulator thin film with 7 quintuple layers and coupled to a cuprate superconductor. In such case with a Fermi energy tuned to be less than $100$meV away from the Dirac point, the minigap of the system can be greater than 0.1meV, which is resolvable in the present experiments. Finally, the oscillatory dependence of the energy of the lowest ABSs can be detected by periodic restoration of the zero bias peak measured through scanning tunneling microscope (STM) when the direction of the external field varies. Note that the STM measurement has a resolution limit, and in the measurement the zero bias peak may occur, while being suppressed, when the splitting of Majorana modes is small. Nevertheless, the maximum height of the zero bias peak is obtained when two Majorana modes are exactly restored in the $2N$ (for a system with $C_{2N}$ symmetry) uniformly distributed angles of the external field, and this phenomenon can reflect the $C_{2N}$ symmetry in the experimental measurement.

In conclusion, we have studied Majorana zero modes protected by 2D cyclic symmetries in the topological crystalline SCs and proposed how to demonstrate the symmetry protection of Majorana modes. A new class of 2D topological superconducting phase has been predicted by inducing $s$-wave superconductivity in the $(001)$-surface of the crystalline insulator Pb$_{1-x}$Sn$_x$Te, with two Majorana modes in each vortex core are protected by the C$_4$ lattice symmetry. Furthermore, we have shown a novel mechanism that with an external field applied in-plane to break the C$_{4}$ symmetry, the two Majorana modes are lifted to finite energy states in general, but, surprisingly, they are restored exactly four times when the field varies $2\pi$, i.e. one cycle in the direction. This phenomenon has a nontrivial correspondence to the four-fold cyclic symmetry of the original system and can provide an unambiguous signature for the lattice-symmetry protection of the MZBMs. We have generalized these results to the system with generic C$_{2N}$ symmetry, in which case there are always two Majorana modes protected by the cyclic lattice symmetry, and the lattice-symmetry protection of such modes can be demonstrated in the $2N$ times of zero-energy crossing when the external field varies one cycle in direction. Our results may provide important insights into the future study of rich physics about symmetry-protection and classification of topological crystalline SCs.

We appreciate the valuable discussions with Jan Zaanen, Jeffrey Teo, Ting-Pong Choy, and Jiansheng Wu. X.J.Liu also thanks  Chen Fang, Zheng-Xin Liu, and Liang Fu for comments. We acknowledge the support from HKRGC through DAG12SC01, Grants No. 602813, No. 605512, and No. HKUST3/CRF/13G.

{\it Note.-}In submitting the initial version of the present manuscript, we note that the result of protection of two Majorana modes by $C_4$ symmetry is also obtained recently in an independent work by C. Fang et al., using a different classification theory~\cite{note2}. Moreover, here we have further proposed a novel scheme to demonstrate the lattice-symmetry protection of MZBMs, and generalized our results to generic $C_{2N}$-symmetric systems.

\noindent

\onecolumngrid

\renewcommand{\thesection}{A-\arabic{section}}
\setcounter{section}{0}  
\renewcommand{\theequation}{A\arabic{equation}}
\setcounter{equation}{0}  
\renewcommand{\thefigure}{A\arabic{figure}}
\setcounter{figure}{0}  

\indent

\section*{\Large\bf Appendix}

\section{Majorana zero modes}

The proximity induced superconductivity in the surface of the topological crystalline superconductor is related to the substrate superconductor, with the induced $s$-wave order given by $\Delta_s=\lambda/(1+\lambda/\Delta_{\rm SC})$~\cite{Sau2010}. Here $\Delta_{\rm SC}$ is the $s$-wave order parameter of the substrate superconductor, and $\lambda$ represents the interface transparency between the substrate and the topological crystalline insulator. For a large interface transparency $\lambda$, the induced $s$-wave order can be close to the magnitude of the order in the substrate.

In the presence of a vortex profile for the $s$-wave order $\Delta_s(r)=\Delta_0(r)e^{in\theta(\bold r)}$, the Hamiltonian reads ${\cal H}_{a_+}(\bold r)=v_F(k_xs_y-k_ys_x)\tau_z-\mu\tau_z+\Delta_0(r)\big[\cos n\theta(\bold r)\tau_x+\sin n\theta(\bold r)\tau_y\bigr]$, with the corresponding Nambu space defined as $\psi^{(+)}=\bigr[c_{a+\uparrow}(\bold r),c_{a+\downarrow}(\bold r),c^\dag_{a+\downarrow}(\bold r),-c^\dag_{a+\uparrow}(\bold r)\bigr]^T$. Due to the self-hermitian property of Majorana fermions, the wave function of a Majorana zero mode generically takes the form $|\gamma\rangle=[\xi_1(\bold r),\xi_2(\bold r),\xi_2^*(\bold r),-\xi_1^*(\bold r)]^T$.
In the polar coordinate the Majorana wave function for ${\cal H}_{a_+}(\bold r)$ satisfies (we rescale $\Delta_0\rightarrow\Delta_0/v_F$ and $\mu\rightarrow\mu/v_F$ for convenience)
\begin{eqnarray}
e^{i\theta}(\partial_r+ir^{-1}\partial_\theta)\xi_1-\Delta_0e^{in\theta}\xi_1^*-\mu\xi_2&=&0,\label{eqn:SIMZBS1}\\
e^{-i\theta}(-\partial_r+ir^{-1}\partial_\theta)\xi_2+\Delta_0e^{in\theta}\xi_2^*-\mu\xi_1&=&0.\label{eqn:SIMZBS2}
\end{eqnarray}
To separate $r$ from $\theta$, we denote by $\xi_1(\bold r)=(u_1+iu_2)e^{i\alpha\theta}$ and $\xi_2(\bold r)=(v_1+iv_2)e^{i\beta\theta}$,
with $\alpha$ and $\beta$ being integers. From the Eqs.~\eqref{eqn:SIMZBS1} and \eqref{eqn:SIMZBS2} we can find that the existence of zero-energy solutions requires $n-\alpha=\alpha+1=\beta$, which gives that $\alpha=(n-1)/2$ and $\beta=(n+1)/2$. Therefore, to have zero modes the winding index of the vortex must be odd integers $n=2m+1$, with which one has that $\alpha=m$ and $\beta=m+1$. We further obtain the differential equations of $u_{1,2}$ and $v_{1,2}$ by
\begin{eqnarray}
\partial_ru_{1,2}-\frac{\alpha}{r}u_{1,2}\mp\Delta_0u_{1,2}-\mu v_{1,2}&=&0,\\
\partial_rv_{1,2}+\frac{\beta}{r}v_{1,2}\mp\Delta_0v_{1,2}+\mu u_{1,2}&=&0,
\end{eqnarray}
where the sign takes ``$-$" for $u_1$ and $v_1$, and takes ``$+$" for $u_2$ and $v_2$.
For convenience we assume that $\Delta_0>0$, and then only $u_2$ and $v_2$ can satisfy the boundary condition which requires that wave function vanishes at $r\rightarrow\infty$. Therefore we must have $u_1=v_1=0$. Denoting by $u_2(r)=e^{-\int_0^rdr'\Delta_0(r')dr'}\tilde{u}_2(r)$ and $v_2(r)=e^{-\int_0^rdr'\Delta_0(r')dr'}\tilde{v}_2(r)$ we get finally
\begin{eqnarray}
(\partial^2_r+\frac{1}{r}\partial_r-\frac{\alpha^2}{r^2})\tilde{u}_2+\mu^2\tilde{u}_2=0,\ \
(\partial^2_r+\frac{1}{r}\partial_r-\frac{\beta^2}{r^2})\tilde{v}_2+\mu^2\tilde{v}_2=0.
\end{eqnarray}
Then the solutions are
\begin{eqnarray}
u_2=e^{-\int_0^rdr'\Delta_0(r')dr'}J_m(\mu r), \ v_2=e^{-\int_0^rdr'\Delta_0(r')dr'}J_{m+1}(\mu r),
\end{eqnarray}
where $J_{m}(r)$ is the $m$-th Bessel function. The Majorana wave function for ${\cal H}_{a_+}(\bold r)$ takes the form (replacing $\Delta_0$ and $\mu$ with $\Delta_0/v_F$ and $\mu/v_F$, respectively)
\begin{eqnarray}
\gamma_a^{(+)}=i\int d^2\bold ru_0(r)\biggr\{J_m(\frac{\mu}{v_F}r)\bigr[c_{a_+\uparrow}(\bold r)- c^\dag_{a_+\uparrow}(\bold r)\bigr]+J_{m+1}(\frac{\mu}{v_F} r)\bigr[c_{a_+\downarrow}(\bold r)e^{i\theta(\bold r)}- c^\dag_{a_+\downarrow}(\bold r)e^{-i\theta(\bold r)}\bigr]\biggr\}e^{im\theta(\bold r)},
\end{eqnarray}
where $u_0(r)=e^{-\int_0^rdr'\frac{\Delta_0(r')}{v_F}dr'}$. Accordingly, the Majorana wave function for ${\cal H}_{a_-}(\bold r)$ is given by
\begin{eqnarray}
\gamma_a^{(-)}=\int d^2\bold ru_0(r)\biggr\{J_m(\frac{\mu}{v_F}r)\bigr[c_{a_-\uparrow}(\bold r)+ c^\dag_{a_-\uparrow}(\bold r)\bigr]+J_{m+1}(\frac{\mu}{v_F} r)\bigr[c_{a_-\downarrow}(\bold r)e^{i\theta(\bold r)}+ c^\dag_{a_-\downarrow}(\bold r)e^{-i\theta(\bold r)}\bigr]\biggr\}e^{im\theta(\bold r)}.
\end{eqnarray}
The other two Majorana zero modes for $H_{b_\pm}$ are obtained by C$_4$ rotation $\gamma^{(\eta)}_b=C_{4}^{(n)-1}\gamma^{(\eta)}_aC_4^{(n)}$ and we have
\begin{eqnarray}
\gamma^{(\pm)}_b=i^{\frac{1\pm1}{2}}\int d^2\bold ru_0(r)\biggr\{J_m(\frac{\mu}{v_F}r)\bigr[c_{b_+\uparrow}(\bold r)\mp c^\dag_{b_+\uparrow}(\bold r)\bigr]+J_{m+1}(\frac{\mu}{v_F} r)\bigr[c_{b_+\downarrow}(\bold r)e^{i\theta(\bold r)}\mp c^\dag_{b_+\downarrow}(\bold r)e^{-i\theta(\bold r)}\bigr]\biggr\}e^{im\theta(\bold r)}.
\end{eqnarray}
Note that the vorticity $n=1$ is relevant for the realistic system, and we shall consider this situation for the following discussion. In this case we have $m=0$.
The transformation relations between the four Majorana zero modes can be summarized by
\begin{eqnarray}\label{eqn:SIMF3}
\gamma^{(\eta)}_b=C_{4}^{(n)^{-1}}\gamma^{(\eta)}_aC_4^{(n)},\ C_{4}^{(n)^{-1}}\gamma^{(\eta)}_bC_4^{(n)}=\eta\gamma^{(\eta)}_a.
\end{eqnarray}
Note the result $C_{2}^{(n)^{-1}}e^{i\theta(\bold r)}C_2^{(n)}=-e^{i\theta(\bold r)}$ has been used. These relations can be reorganized in a more transparent way by redefining that $\gamma_{1,2}=(\gamma^{(+)}_{a,b}+\gamma^{(-)}_{a,b})/\sqrt{2}$ and $\gamma_{3,4}=(\gamma^{(+)}_{a,b}-\gamma^{(-)}_{a,b})/\sqrt{2}$. Then we have
\begin{eqnarray}\label{eqn:SIMF4}
\gamma_{j+1}=C_{4}^{(n)^{-1}}\gamma_{j}C_4^{(n)},\ \gamma_1=C_{4}^{(n)^{-1}}\gamma_{4}C_4^{(n)}.
\end{eqnarray}

\section{Disorder effects and symmetry protection}

The Eq.~\eqref{eqn:SIMF4} reflects the cyclic properties of the four Majorana modes, with which we know that without breaking C$_4$ symmetry the only possible perturbation takes the following form $V_{\rm pert}=\Gamma_0(i\gamma_1\gamma_2+i\gamma_2\gamma_3+i\gamma_3\gamma_4+i\gamma_4\gamma_1)=i2\Gamma_0\gamma^{(-)}_a\gamma^{(-)}_b$. Thus an infinitesimal perturbation without breaking C$_4$ symmetry can gap out $\gamma^{(-)}_{a,b}$. However, the remaining two Majorana zero modes $\gamma^{(+)}_a$ and $\gamma^{(+)}_b$ can be protected by the C$_4$ symmetry. Furthermore, the four-operator interaction $\gamma_1\gamma_2\gamma_3\gamma_4$ cannot be induced by the electron-electron interactions, since this term is forbidden by the C$_4$ symmetry. Therefore, the four-fold cyclic symmetry group protects only two Majorana zero modes, which tells that the topological crystalline superconductor with C$_4$ symmetry is classified by a $Z_2$ invariant. Actually, this $Z_2$ invariant can be defined as the even and odd number of sets of Dirac cones, with each set of them satisfying C$_{4}$ symmetry. If the number of sets is odd, we have two Majorana zero modes protected, while if it is even, no Majorana zero modes can be protected.

\begin{figure}[t]
\includegraphics[width=0.7\columnwidth]{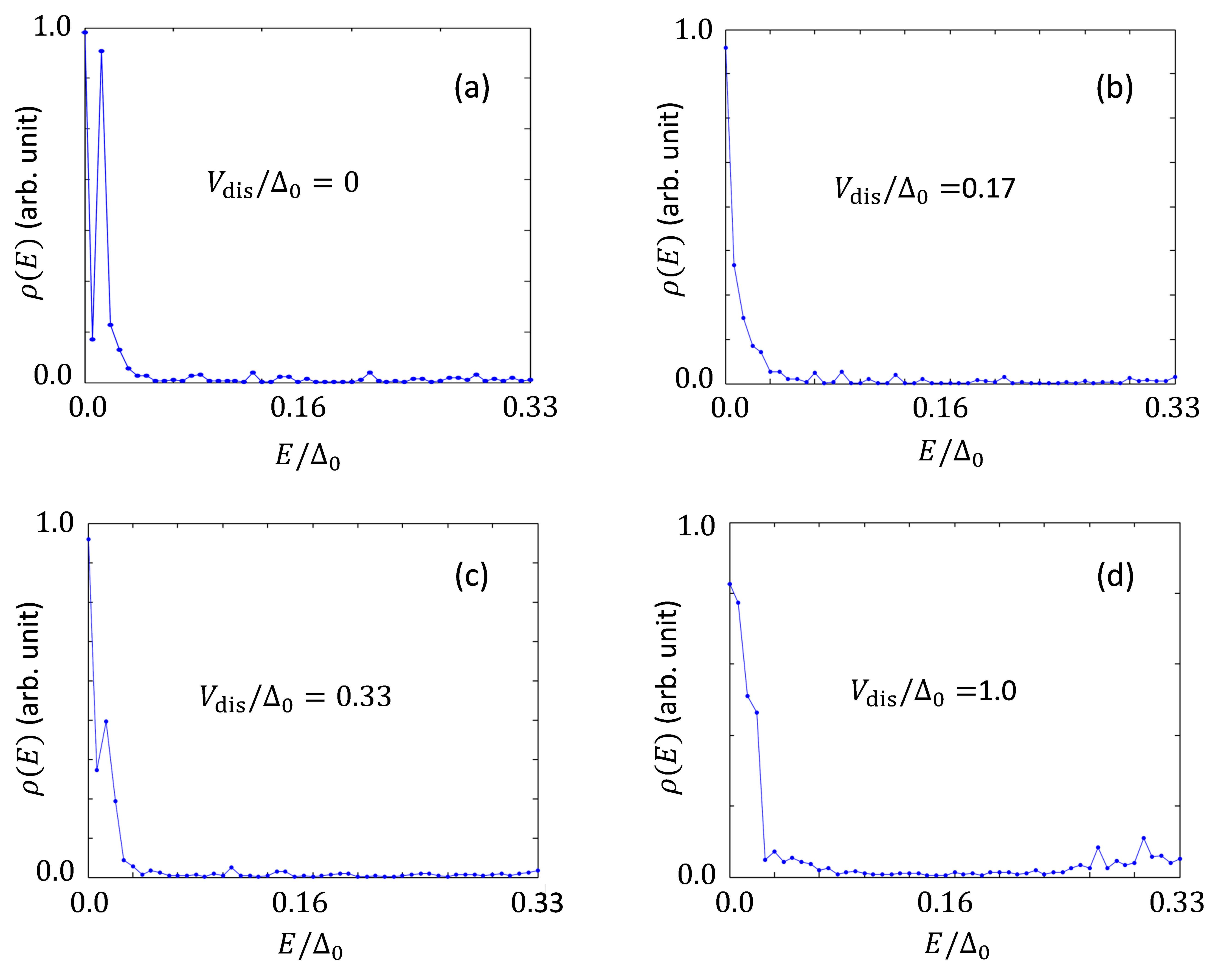}\caption{(Color online)
Density of states $\rho(E)$ in a vortex core with different amplitudes of the random disorder scattering $V_{\rm dis}$. The peak at zero energy represents the Majorana zero bound modes localized in the vortex core, and this peak explicitly exists from the regime without disorder scattering to the one with a relatively strong disorder scattering. (a) $V_{\rm dis}/\Delta_0=0$; (b) $V_{\rm dis}/\Delta_0=0.17$; (c) $V_{\rm dis}/\Delta_0=0.33$; (d) $V_{\rm dis}/\Delta_0=1$. Other parameters in the tight-binding Hamiltonian are rescaled to be dimensionless and chosen that $\varepsilon_0=3.2$, $t_{\rm so}=2.0$, $\mu$=-0.66, and $\Delta_0=0.3$.}
\label{disorder}
\end{figure}

In the realistic topological crystalline superconductor, the defects or disorders generically break the four-fold rotational symmetry. However, the two Majorana modes $\gamma^{(+)}_{a,b}$ can still be protected if by average the disorder respects the four-fold symmetry within the localization length scale of the Majorana modes. Under this condition it can be verified that the hybridization between $\gamma_{a}^{(+)}$ and $\gamma^{(+)}_b$ vanishes, and the C$_4$ symmetry by average provides the protection of the two Majorana modes. This result can be numerically confirmed, as shown in Fig.~\ref{disorder}. In the numerical simulation, we use the tight-binding mode for the topological crystalline insulator Pb$_{1-x}$Sn$_x$Te, as given in Ref.~\cite{Fu3}, and consider the proximity induced $s$-wave superconducting order with the vortex profile $\Delta_s(\bold r)=\Delta_0e^{i\theta(\bold r)}$. For simplicity we approximate $\Delta_0$ to be a constant. The Hamiltonian is written on the three $p$-orbitals of Sn and Te atoms and is given by
\begin{eqnarray}\label{eqn:SIcoupling1}
H_{\rm TB}&=&\varepsilon_0\sum_{j}(-1)^j\sum_{\bold r,s}{\bold p}^\dag_{js}(\bold r)\cdot{\bold p}_{js}(\bold r)-\mu\sum_{j}\sum_{\bold r,s}{\bold p}^\dag_{js}(\bold r)\cdot{\bold p}_{js}(\bold r),\nonumber\\
&&+\sum_{j,j'}t_{jj'}\sum_{\langle\bold r,\bold r'\rangle,s}{\bold p}^\dag_{js}(\bold r)\cdot\hat d_{\bold r\bold r'}\cdot\hat d_{\bold r\bold r'}\cdot{\bold p}_{js}(\bold r')+h.c.\nonumber\\
&&+it_{\rm so}\sum_{j}\sum_{\bold r,s,s'}{\bold p}^\dag_{js}(\bold r)\times{\bold p}_{js'}(\bold r)\cdot\vec \sigma_{ss'}+\sum_{j,\bold r}\bigr[\Delta_s(\bold r){\bold p}_{j\uparrow}(\bold r)\cdot{\bold p}_{j\downarrow}(\bold r)+h.c.\bigr],
\end{eqnarray}
where $j=1,2$ labels the Sn or Te atom, $\bold r$ represents the site, $\bold p_{js}$ and $\bold p^\dag_{js}$ are the annihilation and creation operators for the three $p$-orbitals with $s=\uparrow,\downarrow$, $\mu$ is the chemical potential, $t_{jj'}$ denotes the nearest-neighbor hopping coefficient, $t_{\rm so}$ represents the spin-orbit coupling coefficient, and the induced $s$-wave pairing $\Delta_s(\bold r)$ exists in the $(001)$-surface.

To study the stability of Majorana zero modes, we consider random disorder scatterings in the lattice Hamiltonian with the disorder strength denoted by $V_{\rm dis}$, and require that the disorder configuration is averagely rotation-invariant in the Majorana localization length scale $l_{MF}\sim v_F/\Delta_0$. Then we numerically calculate the density of states $\rho(E)$, whose peak at zero energy $E=0$ represents the Majorana zero modes in the vortex core. From results in Fig.~\ref{disorder} we can find that the peak of $\rho(E)$ at zero energy is explicitly obtained from the regime without disorder scattering to the one with strong disorder scattering ($V_{\rm dis}\sim\Delta_0$). We have also checked many other parameter regimes, and confirmed that the peak of $\rho(E)$ at zero energy can generically be obtained when the disorder scattering respects the C$_4$ symmetry by average in the localization length scale of Majorana modes. These results show that Majorana modes are stable against disorder scattering that is averagely of the C$_4$ symmetry, consistent with the previous analysis.

\section{Demonstrating the symmetry-protection of Majorana modes}

In the presence of mass terms for the Dirac cones in the form $m_j=\alpha(\bold u\times\bold K_j)\cdot\hat e_z$ or $m_j=\alpha(\bold u\cdot\bold K_j)$, where $\bold u$ represents the direction of the displacement vector due to distortion or of the applied external field, the masses $m_1=-m_3=\tilde m_0\cos\phi$ and $m_2=-m_4=\tilde m_0\sin\phi$ for the four Dirac cones, with $\phi$ the angle of the external field. In the bases of $c_{a\pm,\bold k}$ and $c_{b\pm,\bold k}$ the mass terms can be written as
\begin{eqnarray}\label{eqn:SIcoupling1}
H_{m}=\sum_{\bold k}\tilde m_0\cos\phi (c_{a+\uparrow,\bold k}^\dag c_{a-\uparrow,\bold k}-c_{a+\downarrow,\bold k}^\dag c_{a-\downarrow,\bold k})+\sum_{\bold k}\tilde m_0\sin\phi (c_{b+\uparrow,\bold k}^\dag c_{b-\uparrow,\bold k}-c_{b+\downarrow,\bold k}^\dag c_{b-\downarrow,\bold k})+h.c.,
\end{eqnarray}
which explicitly lead to couplings between the Majorana zero modes $\gamma^{(+)}_{a,b}$ and $\gamma^{(-)}_{a,b}$. On the other hand, since the C$_4$ symmetry is broken, an additional $\Gamma_1$-coupling term between $\gamma^{(+)}_a$ and $\gamma^{(+)}_b$ should now be allowed (while we emphasize that the fundamental physics obtained in the present work are independent of the details of the coupling Hamiltonian). Therefore, the total coupling Hamiltonian can be obtained by
\begin{eqnarray}\label{eqn:SIcoupling2}
V_{\rm c}(\phi)=i2\Gamma_0\gamma^{(-)}_a\gamma^{(-)}_b+i2\Gamma_1\sin(2\phi+\beta)\gamma^{(+)}_a\gamma^{(+)}_b +\langle\gamma^{(+)}_a|H_m|\gamma^{(-)}_a\rangle\gamma^{(+)}_a\gamma^{(-)}_a
+\langle\gamma^{(+)}_b|H_m|\gamma^{(-)}_b\rangle\gamma^{(+)}_b\gamma^{(-)}_b.
\end{eqnarray}
Substituting the Eq.~\eqref{eqn:SIcoupling1} and wave-functions of $\gamma_{a,b}^{(\pm)}$ into the above formula we obtain that
\begin{eqnarray}
\langle\gamma^{(+)}_a|H_m|\gamma^{(-)}_a\rangle=i2m_0\cos\phi, \ \
\langle\gamma^{(+)}_b|H_m|\gamma^{(-)}_b\rangle=i2m_0\sin\phi,
\end{eqnarray}
and therefore we finally reach
\begin{eqnarray}\label{eqn:SIcoupling3}
V_{\rm c}(\phi)=i2\Gamma_0\gamma^{(-)}_a\gamma^{(-)}_b+i2\Gamma_1\sin(2\phi+\beta)\gamma^{(+)}_a\gamma^{(+)}_b+ i2m_0\cos\phi\gamma^{(+)}_a\gamma^{(-)}_a+i2m_0\sin\phi\gamma^{(+)}_b\gamma^{(-)}_b,
\end{eqnarray}
where the coupling amplitude reads $m_0=2\tilde m_0\int dr\pi ru^2_0(r)\bigr[J_0^2(\frac{\mu}{v_F}r)-J_1^2(\frac{\mu}{v_F}r)\bigr]$.
Note that the variation of $\phi$ by $\pi/2$ is equivalent to do a $C_4^{(n)}(\pi/2)$ rotation, which transforms $\gamma^{(+)}_a\leftrightarrow\gamma^{(+)}_b$ and changes the sign of the coupling coefficient. Therefore $\Gamma_1$-coupling term must be proportional to $\sin(2\phi+\beta)$, with $\beta$ an arbitrary constant which can be material-dependent. In general the perturbation by an in-plane vector- or pseudovector-type field always induces the symmetry-breaking couplings as given in Eq.~\eqref{eqn:SIcoupling3}.

Solving the spectra of the coupling Hamiltonian~\eqref{eqn:SIcoupling3} we can find that four zero-energy crossings are obtained upon varying the direction angle $\phi$ of the external field by $4\pi$. This phenomenon can be understood in an intuitive picture as introduced below. In the coupling Hamiltonian $V_{\rm c}(\phi)$ the C$_4$ symmetry is broken for any fixed $\phi$. However, if we take the external field and the topological crystalline superconductor as a whole, the entire system is invariant under the C$_4$ rotation. This implies that we generically have
\begin{eqnarray}\label{eqn:SItransformation}
V_{\rm c}(\phi-\pi/2)=C_4^{(n)^{-1}}V_{\rm c}(\phi)C_4^{(n)}.
\end{eqnarray}
On the other hand, the C$_4$ rotation transforms that $\gamma_a^{(\pm)}\rightarrow\gamma_b^{(\pm)}$ and $\gamma_b^{(\pm)}\rightarrow\pm\gamma_a^{(\pm)}$. Then, for the four Majorana zero bound modes which form two complex fermions (Andreev Bound states) as $f_{a,b}=\gamma_{a,b}^{(+)}+i\gamma_{a,b}^{(-)}$, the variation of $\pi/2$ in $\phi$ equivalently transforms $f_a\rightarrow f_b$ and $f_b\rightarrow f_a^\dag$. This result shows that the ground state of the system changes fermion parity, and a zero-energy crossing occurs when $\phi$ advances $\pi/2$. Accordingly, if varying $\phi$ by $2\pi$, i.e. one circle, we get $4$ times of restoration of the Majorana zero bound states. This novel phenomenon can provide an unambiguous signature for the lattice-symmetry protection of the Majorana modes. The formula~\eqref{eqn:SItransformation} can be directly generalized to the topological crystalline superconductor with generic C$_{2N}$ symmetry, yielding
\begin{eqnarray}
V_{\rm c}(\phi-\pi/2)=C_{2N}^{(n)^{-1}}V_{\rm c}(\phi)C_{2N}^{(n)}.
\end{eqnarray}
Then varying $\phi$ by $2\pi$ leads to $2N$ times of restoration of Majorana zero modes.



\noindent


\begin{thebibliography}{99}
\bibitem{note1} In this case the term $2f^\dag f-1$ is allowed, while the coupling $i\gamma_+\gamma_-$ is forbidden by the $C_4^{(n)}$ symmetry.
\end{thebibliography}
\end{document}